\documentclass[lettersize,journal]{IEEEtran}
\usepackage{amsmath,amsfonts}
\usepackage{algorithmic}
\usepackage{algorithm}
\usepackage{array}
\usepackage[caption=false,font=normalsize,labelfont=sf,textfont=sf]{subfig}
\usepackage{textcomp}
\usepackage{stfloats}
\usepackage{url}
\usepackage{verbatim}
\usepackage{graphicx}
\usepackage{stfloats}
\usepackage{cite}
\usepackage{amssymb}
\usepackage[table]{xcolor}
\definecolor{lightergreen}{rgb}{0.5, 0.8, 0.4} % Adjust RGB values to your preference
\definecolor{lighterorange}{rgb}{1, 0.8, 0.7}

\newcommand{\LOOPJCCP}{$\mathcal{LOOP-JCCP}$}
\newcommand{\LOOPLCP}{$\mathcal{LOOP-LC}\hspace{.1cm}2.0$}
\begin{document}

\title{Learning to Optimize Joint Chance-constrained Power Dispatch Problems}

\author{Meiyi Li,~\IEEEmembership{Student Member,~IEEE}, Javad Mohammadi,~\IEEEmembership{Senior Member,~IEEE}

        % <-this % stops a space

\thanks{Meiyi Li and Javad Mohammadi are with the Department of Civil, Architectural, and Environmental Engineering, 
The University of Texas at Austin, Austin, USA
 (e-mail: \{meiyil, javadm\}@utexas.edu).
% Manuscript received April 19, 2021; revised August 16, 2021.
}
}

\markboth{Accepted by CSEE Journal of Power and Energy Systems in Dec. 2024}%
{Shell \MakeLowercase{\textit{et al.}}: A Sample Article Using IEEEtran.cls for IEEE Journals}


\maketitle

\begin{abstract} \color{black}
The ever-increasing integration of stochastic renewable energy sources into power systems operation is making the supply-demand balance more challenging. While joint chance-constrained methods are equipped to model these complexities and uncertainties, solving these models using the traditional iterative solvers is time-consuming and can hinder real-time implementation.
To overcome the shortcomings of today's solvers, we propose a fast, scalable, and explainable machine learning-based optimization proxy. Our solution, called Learning to Optimize the Optimization of Joint Chance-Constrained Problems (\LOOPJCCP), is iteration-free and solves the underlying problem in a single-shot. Our model uses a polyhedral reformulation of the original problem to manage constraint violations and ensure solution feasibility across various scenarios through customizable probability settings. To this end, we build on our recent deterministic solution (\LOOPLCP) by incorporating a set aggregator module to handle uncertain sample sets of varying sizes and complexities. Our results verify the feasibility of our near-optimal solutions for joint chance-constrained power dispatch scenarios. Additionally, our feasibility guarantees increase the transparency and interpretability of our method, which is essential for operators to trust the outcomes. We showcase the effectiveness of our model in solving the stochastic energy management problem of Virtual Power Plants (VPPs). Our numerical findings complement our theoretical justifications and demonstrate great flexibility in parameter tuning, adaptability to diverse datasets, and increased computational speed. 
\end{abstract}

\vspace{.1cm}

\section{Introduction}
\IEEEPARstart{T}{he} transition to sustainable energy, with its emphasis on integrating renewable sources, introduces increased uncertainties in power dispatch challenges \cite{navidi2023coordinating}. This shift puts pressure on reserve capacities and power flows, posing risks to the resilience of electric networks. Consequently, there is a critical need to develop swift and effective solutions for stochastic power dispatch problems\cite{roald2023power}, \cite{geng2019data}. 

{\color{black}Joint chance-constrained methods, which consider the collective probability of multiple constraints, provide solutions that are appropriately balanced, ensuring both security and cost-effectiveness at the system level \cite{baker2019joint, hou2022data, wang2024risk}. These methods are extensively applied in various power dispatch scenarios, including frequency security \cite{tian2024joint, yang2024stochastic}, scenario generation \cite{zhang2024efficient}, load forecasting \cite{zhou2024addressing}, and the coordination of electricity and heating systems \cite{zhai2024data}.}

{\color{black}Data-driven techniques are typically used for joint chance-constrained problems to achieve a balance between feasibility and cost optimality. }The scenario approach, a common method, involves finding solutions that remain feasible across all historical scenarios, offering probabilistic feasibility guarantees but often resulting in overly conservative outcomes with higher costs due to large sample sizes and dimensionality challenges \cite{calafiore2006scenario, hou2022data, pena2020dc}. Another technique, the conditional value-at-risk (CVaR) approximation, aims to quantify expected losses in the worst-case scenarios. However, its sample approximation can suffer from adjustability and dimensionality issues, potentially leading to unsafe risk estimations \cite{geng2019data, roald2023power}. Additionally, robust optimization, which ensures probabilistic constraints within a predefined uncertainty set based on statistical moments, faces difficulties in selecting an optimal safety parameter to balance coverage of probability mass and cost reduction \cite{roald2015security, hou2021data}. {\color{black} 
In addition to moment-based sets, the Wasserstein distance has also been applied to measure distributional uncertainty \cite{zhou2024distributionally,yang2024low}. The Wasserstein distance quantifies the difference between empirical data distributions and the true underlying distribution. %In contrast, our moment-based Polyhedron Reformulation uses statistical moments, specifically the average and maximum values of the input features, to determine the bounds of the uncertainty set.
%Besides the type of moment-based set, Wasserstein distance has also been applied to measure the distributional uncertainty \cite{zhou2024distributionally,yang2024low}. Wasserstein distance quantifies the difference between empirical data distributions and the true underlying distribution, allowing for robust decision-making under ambiguity.
}

% Data-driven techniques are typically used for joint chance-constrained problems to achieve a balance between feasibility and cost optimality. The scenario approach, a common method, involves finding solutions that remain feasible across all historical scenarios, offering probabilistic feasibility guarantees but often resulting in overly conservative outcomes with higher costs due to large sample sizes and dimensionality challenges \cite{calafiore2006scenario, hou2022data, pena2020dc}. Another technique, the conditional value-at-risk (CVaR) approximation, aims to quantify expected losses in the worst-case scenarios. However, its sample approximation can suffer from adjustability and dimensionality issues, potentially leading to unsafe risk estimations \cite{geng2019data, roald2023power}. Additionally, robust optimization, which ensures probabilistic constraints within a predefined ellipsoidal uncertainty set based on statistical moments, faces difficulties in selecting an optimal safety parameter to balance coverage of probability mass and cost reduction \cite{roald2015security, hou2021data}.

{\color{black}Joint chance-constrained methods introduce greater computational challenges due to complex dependencies among constraints, which significantly increase computational demands.} To enhance efficiency in handling these challenges, Machine Learning (ML) has been increasingly utilized across various power system applications, particularly in tasks that require repetitive execution \cite{fioretto2020predicting,pan2020deepopf,biagioni2020learning, li2023learningADMM,li2023machine,keerthisinghe2018energy}. Furthermore, the integration of Explainable AI (XAI) techniques not only improves the transparency of ML models but also builds trust and provides clear insights into decision-making processes. This advancement allows stakeholders to better understand, validate, and effectively implement ML solutions, thereby enhancing the robustness and accountability of power system operations.

{\color{black}Existing ML-assisted methods for chance-constrained optimization have predominantly focused on probabilistic prediction under uncertainty \cite{dalal2019chance, ning2021deep, liang2024joint} or status classification \cite{baker2019joint} to aid solvers in the subsequent solution search. However, the iterative nature of traditional solvers often limits their utility in real-time decision-making \cite{li2023learning}, and these studies still largely depend on such solvers for the optimization process \cite{liang2024joint}. This reliance highlights a significant research gap -- the need for tractable surrogate models that can directly predict optimal solutions for chance-constrained problems.} The potential of learning-based methods to overcome the computational challenges inherent to chance-constrained methods remains largely untapped and represents a substantial opportunity for advancement.

In this paper, we develop a neural approximator to predict the optimal solution for joint chance-constrained power dispatch problems, specifically focusing on Virtual Power Plants (VPPs). Our approach, called Learning to Optimize the Optimization of Joint Chance-Constrained
Problems (\LOOPJCCP), builds upon previous efforts, addressing the unique demands of joint chance-constrained problems:

\begin{itemize}
    \item \underline{\textit{Near-Optimality}}: Our neural approximator is designed to efficiently map input parameters and sample sets to high quality solutions (that are close to optimal outputs of joint chance-constrained power dispatch problems). 
    \item \underline{\textit{Constraint Compliance with Adjustable Probability}}: Our model features a closed-form, explainable feasibility module with customizable probability settings, offering flexibility during both offline training and online testing. This module ensures predicted solutions consistently adhere to necessary constraints while providing transparency and interpretability in how probability settings impact solution feasibility, aligning with XAI principles.
    \item \underline{\textit{Rapid and Repetitive Execution}}: {\color{black}Using an iteration-free structure, our method is multiple times faster than traditional solver-based models.} The proposed \LOOPJCCP~model substantially enhances the efficiency of timely and repetitive tasks. This speed advantage is key to its effectiveness in dynamic operational environments and applications with moving time horizons.
    \item \underline{\textit{Scalability to Different Input Sequences and Scales}}: Demonstrating remarkable adaptability, the \LOOPJCCP~model's structure is uniquely designed to be insensitive to the order of input samples. It can effortlessly handle varying numbers of samples, showcasing its robust scalability and flexibility across different operational scales.
\end{itemize}

Central to our approach is a novel polyhedron reformulation of the original chance-constrained problem, enabling the model to effectively learn the constraints specific to these problems. We also incorporate insights from our recent work on Learning to Optimize Linearly Constrained Problems version 2.0 \cite{li2023toward} (\LOOPLCP), facilitating adjustments in violation rates. These innovations collectively enhance the performance of the proposed \LOOPJCCP~model in terms of cost optimality, speed, feasibility, and scalability.
 
\section{Chance-constrained power dispatch problem formulation}

\subsection{Problem Formulation}
\begin{figure}[htbp]
\centering
\setlength{\abovecaptionskip}{0.cm}
\includegraphics[width=1\columnwidth]{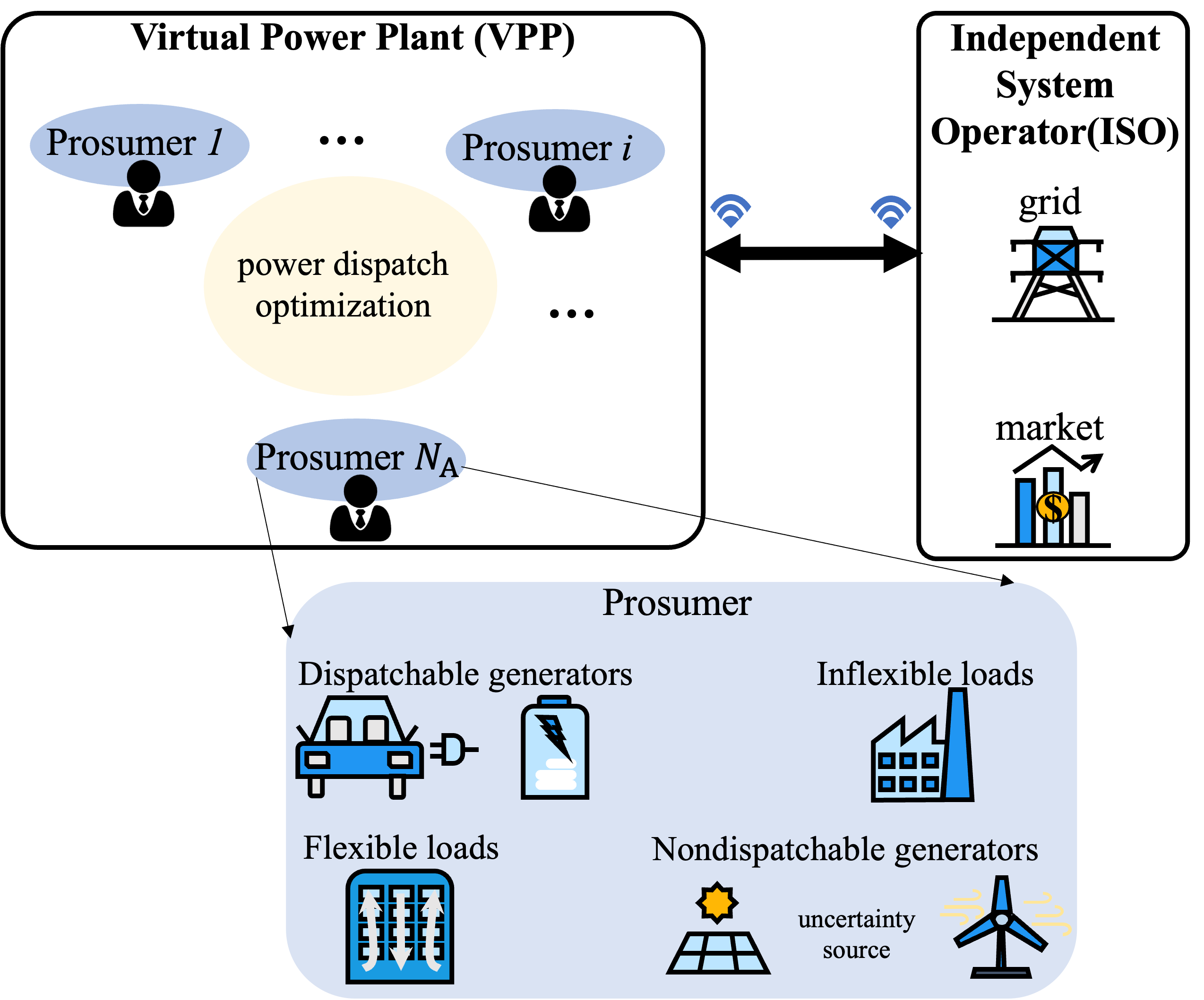}
\caption{Diagram of the VPP in this study, highlighting uncertainties from non-dispatchable sources like photovoltaic arrays and wind turbines, due to variations between actual and predicted renewable energy outputs.}
\centering
\label{f:vpp}
\end{figure}

Consider a VPP comprising of \( N_{\texttt{A}} \) prosumers (producers $+$ consumers), each represented by an index \( i \) where \( i = 1, \ldots, N_{\texttt{A}} \). As illustrated in Fig \ref{f:vpp}, each prosumer owns a diverse range of DERs. This includes dispatchable generators such as energy storage systems and electric vehicles, as well as non-dispatchable generators like photovoltaic arrays and wind turbines. The aggregation also encompasses both inflexible (critical) and flexible (noncritical) loads. These prosumers may be associated with various distribution utilities. The overarching goal of the VPP is to effectively coordinate these prosumers, ensuring optimal energy contribution to the grid under uncertainty of renewable energy while simultaneously optimizing the individual utility functions of each prosumer.

The power dispatch problem for the VPP is formulated as presented in \eqref{power dispatch}. Equations \eqref{generator} and \eqref{load} specify that the dispatchable generator \( P_{\texttt{G}}^{i} \) and the flexible load \( P_{\texttt{L}}^{i} \) must be maintained within their respective safe operating ranges. The overall output power of an prosumer is defined by \eqref{output def} as \( P_{\texttt{o}}^{i} \), where \( P_{\texttt{NG}}^{i} \) represents the power from non-dispatchable generators (i.e., predicted power from renewable sources), and \( P_{\texttt{IL}}^{i} \) represents the power of inflexible loads. Local distribution utility constraints are accounted for in  \eqref{output limit}. Additionally, \eqref{balance} ensures that the total output of the VPP is in accordance with the planned production schedule \( P_{\texttt{Sch}} \). The objective function, defined in \eqref{object}, assigns a quadratic utility function to each dispatchable generator and flexible load, following the formulation described in \cite{hug2015consensus+}.

\begin{subequations}
\label{power dispatch}
\begin{gather}
\min f=\sum_{i=1}^{N_{\texttt{A}}}\left((\beta_{\texttt{G1}}^{i} {P_{\texttt{G}}^{i}}^2+ \beta_{\texttt{G2}}^{i} {P_{\texttt{G}}^{i}}) +  (\beta_{\texttt{L1}}^{i} {P_{\texttt{L}}^{i}}^2+ \beta_{\texttt{L2}}^{i} {P_{\texttt{L}}^{i}})    \right)\label{object}\\
    P_{\texttt{Gmin}}^{i}\leq P_{\texttt{G}}^{i} \leq P_{\texttt{Gmax}}^{i}, i=1,...,N_{\texttt{A}}\label{generator}\\
    P_{\texttt{Lmin}}^{i}\leq P_{\texttt{L}}^{i} \leq P_{\texttt{Lmax}}^{i}, i=1,...,N_{\texttt{A}}\label{load}\\
    P_{\texttt{o}}^{i}=P_{\texttt{G}}^{i}-P_{\texttt{L}}^{i}+P_{\texttt{NG}}^{i}-P_{\texttt{IL}}^{i}\label{output def}\\
    P_{\texttt{omin}}^{i}\leq P_{\texttt{o}}^{i}\leq P_{\texttt{omax}}^{i}\label{output limit}\\
    \sum_{i=1}^{N_{\texttt{A}}}P_{\texttt{o}}^{i}=P_{\texttt{Sch}}\label{balance}
\end{gather}
\end{subequations}

We account for the uncertainty \( \epsilon^{i} \) arising from non-dispatchable generators (renewable generators) associated with each prosumer \( i \), where \( i = 1, \ldots, N_{\texttt{A}} \). This uncertainty results in deviations from the predicted power \( P_{\texttt{NG}}^{i} \) to \( P_{\texttt{NG}}^{i} + \epsilon^{i} \). Consequently, we model the responsive adjustment of dispatchable generators from \( P_{\texttt{G}}^{i} \) to \( P_{\texttt{G}}^{i} - \alpha_{\texttt{G}}^i \sum_{i=1}^{N_{\texttt{A}}} \epsilon^{i} \) and flexible loads from \( P_{\texttt{L}}^{i} \) to \( P_{\texttt{L}}^{i} + \alpha_{\texttt{L}}^i \sum_{i=1}^{N_{\texttt{A}}} \epsilon^{i} \). These adjustments will help satisfy \eqref{balance} in the face of uncertainties. Here, \( \alpha_{\texttt{G}}^i \) and \( \alpha_{\texttt{L}}^i \) represent the participation factors for dispatchable generators and flexible loads, respectively. These factors are calculated as:

\begin{gather} \alpha_{\texttt{G}}^i=\frac{P_{\texttt{Gmax}}^{i}}{\sum_{i=1}^{N_{\texttt{A}}} (P_{\texttt{Gmax}}^{i}+P_{\texttt{Lmax}}^{i})},
    \alpha_{\texttt{L}}^i=\frac{P_{\texttt{Lmax}}^{i}}{\sum_{i=1}^{N_{\texttt{A}}} (P_{\texttt{Gmax}}^{i}+P_{\texttt{Lmax}}^{i})}
\end{gather}
\normalsize

Subsequently, the deterministic power dispatch problem in \eqref{power dispatch} is restructured into the chance-constrained model in \eqref{power dispatch cc}, substituting \( P_{\texttt{o}}^{i} \) for simplicity.
In \eqref{power dispatch cc}, the joint chance constraint \eqref{joint constraint} ensures that all individual constraints are satisfied simultaneously with a probability greater than \( 1-\varepsilon \). The parameter \( \varepsilon \) (ranging from 0 to 1) sets the acceptable level of violation within the chance constraint framework.

\begin{figure*}[htb]

\begin{subequations}
\label{power dispatch cc}
\begin{gather}
\min f=\sum_{i=1}^{N_{\texttt{A}}}\left((\beta_{\texttt{G1}}^{i} {P_{\texttt{G}}^{i}}^2+ \beta_{\texttt{G2}}^{i} {P_{\texttt{G}}^{i}}) +  (\beta_{\texttt{L1}}^{i} {P_{\texttt{L}}^{i}}^2+ \beta_{\texttt{L2}}^{i} {P_{\texttt{L}}^{i}})    \right)\label{object cc}\\
\mathbb{P}\begin{pmatrix}
P_{\texttt{Gmin}}^{i}\leq P_{\texttt{G}}^{i}-\alpha_{\texttt{G}}^i \sum_{i=1}^{N_{\texttt{A}}}\epsilon^{i} \leq P_{\texttt{Gmax}}^{i}, i=1,...,N_{\texttt{A}}\\
P_{\texttt{Lmin}}^{i}\leq P_{\texttt{L}}^{i} -\alpha_{\texttt{L}}^i \sum_{i=1}^{N_{\texttt{A}}}\epsilon^{i}\leq P_{\texttt{Lmax}}^{i}, i=1,...,N_{\texttt{A}}\\
P_{\texttt{omin}}^{i}\leq (P_{\texttt{G}}^{i}-\alpha_{\texttt{G}}^i  \sum_{i=1}^{N_{\texttt{A}}}\epsilon^{i})+(P_{\texttt{NG}}^{i}+\epsilon^{i})+P_{\texttt{NG}}^{i}-(P_{\texttt{L}}^{i}+\alpha_{\texttt{L}}^i \sum_{i=1}^{N_{\texttt{A}}}\epsilon^{i})-P_{\texttt{IL}}^{i} \leq P_{\texttt{omax}}^{i}, i=1,...,N_{\texttt{A}}
\end{pmatrix}\geq 1-\varepsilon  \label{joint constraint}  \\
\sum_{i=1}^{N_{\texttt{A}}}(P_{\texttt{G}}^{i}+P_{\texttt{NG}}^{i}-P_{\texttt{L}}^{i}-P_{\texttt{IL}}^{i})=P_{\texttt{Sch}}\label{balance cc}
\end{gather}
\end{subequations}
\normalsize
\end{figure*}

Let vector \( \boldsymbol{\epsilon} = [\epsilon^{i} \mid i = 1, \ldots, N_{\texttt{A}}] \) represent the collection of all uncertainties within the VPP. The chance-constrained power dispatch problem, expressed in its compact form, is presented in \eqref{power dispatch cc compact}. In this formulation, \( \mathbf{x} = [P_{\texttt{Sch}}, P_{\texttt{NG}}^{i}, P_{\texttt{IL}}^{i} \mid i = 1, \ldots, N_{\texttt{A}}] \) denotes the collection of input parameters, while \( \mathbf{u} = [P_{\texttt{G}}^{i}, P_{\texttt{L}}^{i} \mid i = 1, \ldots, N_{\texttt{A}}] \) represents the stacked optimization variables. The matrices \( \mathbf{A}_{\texttt{eq}}, \mathbf{B}_{\texttt{eq}}, \mathbf{A}_{\texttt{ineq}}, \mathbf{B}_{\texttt{ineq}}, \mathbf{C}_{\texttt{ineq}} \), and vector \( \mathbf{b}_{\texttt{ineq}} \) constitute the compact parameters.

\begin{subequations}
\label{power dispatch cc compact}
\begin{gather}
    \min f(\mathbf{u}) \label{objective function power dispatch cc compact} \\
\texttt{s.t.}~~~\mathbf{A}_{\texttt{eq}}\mathbf{u}+\mathbf{B}_{\texttt{eq}}\mathbf{x}=\textbf{0}\label{power dispatch eq compact}\\
\mathbb{P}(\mathbf{A}_{\texttt{ineq}}\mathbf{u}+\mathbf{B}_{\texttt{ineq}}\mathbf{x}+\mathbf{C}_{\texttt{ineq}}\boldsymbol{\epsilon}+\mathbf{b}_{\texttt{ineq}}\leq\textbf{0})\geq 1-\varepsilon 
\label{power dispatch ineq compact}
\end{gather}
\end{subequations}
\normalsize

In order to convert probabilistic constraints into deterministic ones, a set of samples (also referred to as scenarios) \( \{\boldsymbol{\epsilon}^{[k]}\}_{k=1}^{N_{\texttt{scen}}} \) is utilized. The VPP faces the challenge of repeatedly solving the problem as defined in \eqref{power dispatch cc compact}, with varying input parameters \( \mathbf{x} \) and sample sets. The objective of this optimization process is to find an optimal solution \( \mathbf{u} \) that minimizes the objective function \( f \) while also satisfying probabilistic guarantees.

\subsection{Overview of Key Reformulation Strategies}

To solve any joint chance-constrained problem, it must first be reformulated into a tractable format. This subsection introduces three widely-used reformulations: (1) the scenario approach; (2) the sample average approximation; and (3) robust optimization-based methods.

\subsubsection{Scenario Approach}
The scenario approach stands out for its simplicity and effectiveness. Its most compelling attribute is its universal applicability. This method (presented by \eqref{eq:sa compact}) is entirely data-driven and does not rely on any assumptions regarding the underlying probability distribution. Essentially, the scenario approach identifies an optimal solution that remains feasible across all \( N_{\texttt{Scen}} \) scenarios.

\begin{subequations}
\label{eq:sa compact}
\begin{gather}
    \min f(\mathbf{u}) \label{eq:sa_objective_function compact} \\
\texttt{s.t.}~~~\mathbf{A}_{\texttt{eq}}\mathbf{u}+\mathbf{B}_{\texttt{eq}}\mathbf{x}=\textbf{0}\label{eq:sa eq compact}\\
\mathbf{A}_{\texttt{ineq}}\mathbf{u}+\mathbf{B}_{\texttt{ineq}}\mathbf{x}+\mathbf{C}_{\texttt{ineq}}\boldsymbol{\epsilon}^{[k]}+\mathbf{b}_{\texttt{ineq}}\leq\textbf{0}, k=1,...,N_{\texttt{Scen}}
\label{eq:sa ineq compact}
\end{gather}
\end{subequations}
\normalsize

The effectiveness of the scenario approach is significantly influenced by the availability of a sample set. After the sample set is established, this approach allows for only limited adjustments to the conservatism of the derived solution. Another major challenge faced by this method is its handling of dimensionality, especially when there is a requirement for a large sample size.

\subsubsection{CVaR Reformulation Using Sample Approximation}
CVaR is a risk measure that estimates the expected loss in the worst percentage of scenarios. Sample approximation is commonly used to approximate CVaR. By introducing an auxiliary variable \( \beta^{[k]} \) for each uncertainty scenario \( k \), where \( k = 1, \ldots, N_{\texttt{Scen}} \), to represent uncertain constraint violations \cite{sun2014asymptotic}, the joint chance-constrained problem in \eqref{power dispatch} can be reformulated into a more tractable convex problem as in \eqref{eq:cvar  compact}.

\begin{subequations}
\label{eq:cvar compact}
\begin{gather}
 \min f(\mathbf{u}) \label{eq:cvar_objective_function compact} \\
\texttt{s.t.}~~~\mathbf{A}_{\texttt{eq}}\mathbf{u}+\mathbf{B}_{\texttt{eq}}\mathbf{x}=\textbf{0}\label{eq:cvar eq compact}\\
\mathbf{A}_{\texttt{ineq}}\mathbf{u}+\mathbf{B}_{\texttt{ineq}}\mathbf{x}+\mathbf{C}_{\texttt{ineq}}\boldsymbol{\epsilon}^{[k]}+\mathbf{b}_{\texttt{ineq}}\leq \beta^{[k]}\textbf{1}, \nonumber\\
k=1,...,N_{\texttt{Scen}}
\label{eq:cvar ineq compact}\\
\frac{1}{N_{\texttt{Scen}}}\sum_{k=1}^{N_{\texttt{Scen}}} \beta^{[k]}-(1-\varepsilon)\beta^0\leq0 \label{eq:cvar_avg}\\
\beta^{[k]}\geq \beta^0, \forall k=1,...,N_{\texttt{Scen}}\label{eq:cvar_limit}   
\end{gather}
\end{subequations}

\normalsize

\noindent where auxiliary variable $\beta^0$ denotes a threshold such that the chance constraint violation exceeds $\beta^0$ in the worst \(\varepsilon\) percentile cases. And constraint \eqref{eq:cvar_avg} and \eqref{eq:cvar_limit} imply $\beta^0\leq0$.

Similar to the scenario approach, the CVaR reformulation using sample approximation offers limited flexibility in adjusting the conservatism of the solution once a sample set is given. This method also faces significant challenges with dimensionality, particularly when a large sample size is required. Furthermore, integrating the CVaR reformulation into a neural network is complex due to the auxiliary variables involved. The number of these variables correlates with the sample number \( N_{\texttt{Scen}} \). Since traditional neural network structures produce outputs of fixed size, mimicking the CVaR reformulation with variable sample sets may pose additional difficulties.

\subsubsection{Robust Optimization with Moment-Based Reformulation}

Another common strategy for addressing chance-constrained problems is robust optimization, where probabilistic constraints must hold across all scenarios within a defined safe approximation set \cite{geng2019data}. This set is often characterized using the first and second moments (mean and covariance) of uncertainty \cite{roald2023power}, leading to a robust formulation as follows:

\begin{subequations}
\label{eq:ro compact}
\begin{gather}
    \min f(\mathbf{u}) \label{eq:ro_objective_function compact} \\
\texttt{s.t.}~~~\mathbf{A}_{\texttt{eq}}\mathbf{u}+\mathbf{B}_{\texttt{eq}}\mathbf{x}=\textbf{0}\label{eq:ro eq compact}\\
\mathbf{A}^{\left \{ r \right \}}_{\texttt{ineq}}\mathbf{u}+\mathbf{B}^{\left \{ r \right \}}_{\texttt{ineq}}\mathbf{x}+\mathbf{C}^{\left \{ r \right \}}_{\texttt{ineq}}\boldsymbol{\epsilon}_{\texttt{avg}}+\mathbf{b}^{\left \{ r \right \}}_{\texttt{ineq}} \nonumber\\
+s(1-\varepsilon)\left \| \mathbf{C}^{\left \{ r \right \}}_{\texttt{ineq}} \boldsymbol{\epsilon}_{\texttt{std}}\right \|_2
\leq 0, \forall r
\label{eq:ro ineq compact}
\end{gather}
\end{subequations}
\normalsize

In this formulation, \( \mathbf{D}^{\left \{ r \right \}} \) represents the \( r \)th row vector in matrix \( \mathbf{D} \). The vectors \( \boldsymbol{\epsilon}_{\texttt{avg}} \) and \( \boldsymbol{\epsilon}_{\texttt{std}} \) denote the mean and standard deviation, respectively, of the sample set \( \{\boldsymbol{\epsilon}^{[k]}\}_{k=1}^{N_{\texttt{Scen}}} \). The safety parameter \( s > 0 \) plays a critical role; a larger value of \( s(1-\varepsilon) \) tightens the constraints, resulting in a smaller feasible set and potentially more expensive solutions.

Integrating moment-based robust reformulation into a neural network poses challenges, as it requires addressing constraints individually—each row in the parameter matrix represents a distinct constraint. This characteristic complicates the integration and parallel operations within a neural network structure.

All three reformulations transform the joint chance-constrained problem into deterministic forms and rely on iterative solvers for optimal solutions (\eqref{power dispatch}). We introduce a neural approximator, \LOOPJCCP, developed to predict solutions for these problems efficiently. Additionally, due to integration challenges with existing methods, we propose a novel polyhedron reformulation designed to improve the neural network's learning efficiency.

\section{Machine leaning method}
\subsection{Overview of the Proposed \LOOPJCCP~Model}
\begin{figure}[htbp]
\centering
\setlength{\abovecaptionskip}{0.cm}
\includegraphics[width=1\columnwidth]{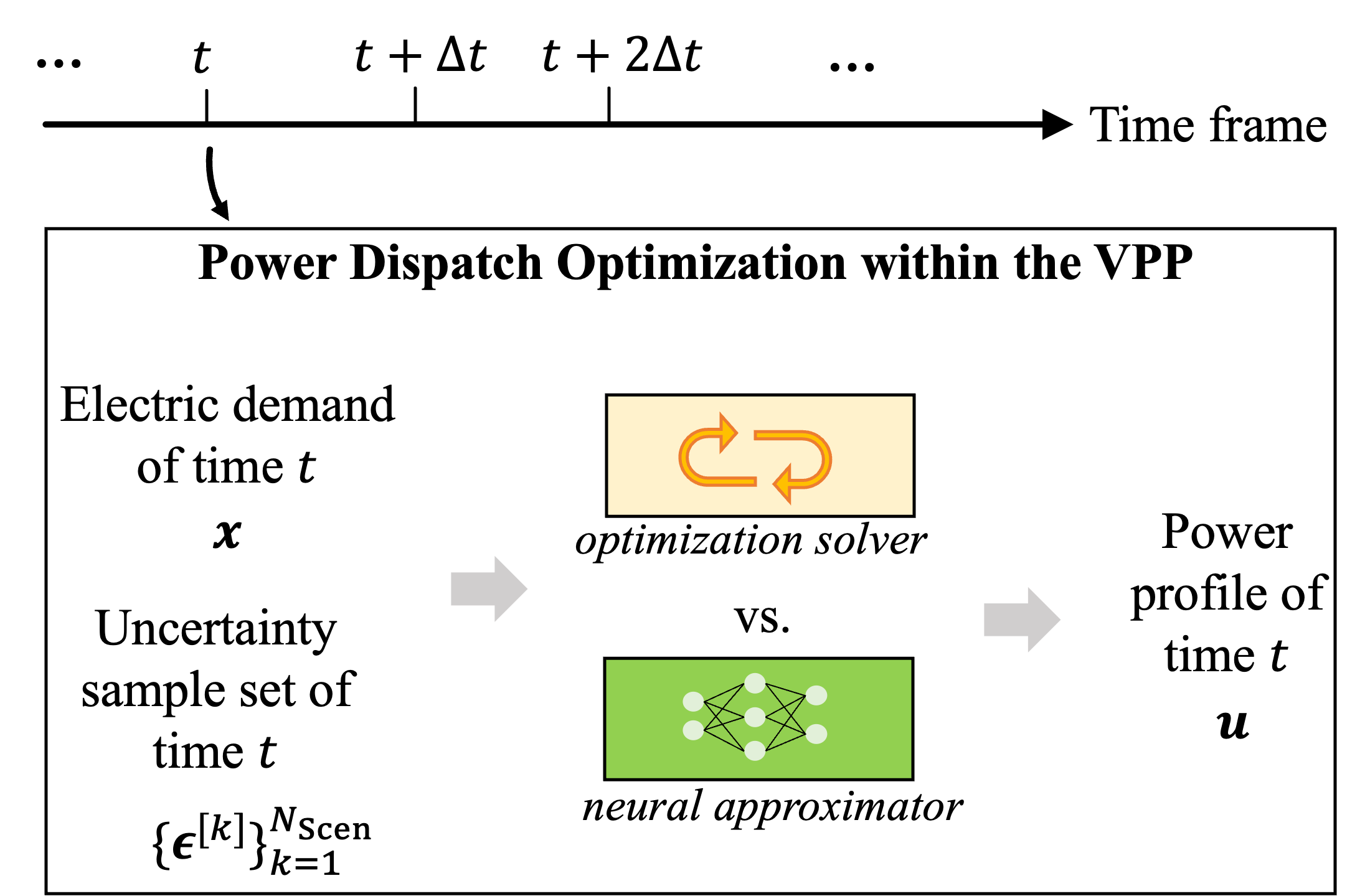}
\caption{\LOOPJCCP~model as an alternative to iterative solvers to deal with the task of repeatedly solving power dispatch optimization for VPPs.}
\centering
\label{f:basic}
%\vspace{-0.1in}
\end{figure}

Our proposed approach seeks to replace traditional solvers with a trained neural network (denoted as \( \xi_{\mathbf{w}} \)) to tackle problem \eqref{power dispatch cc compact}. This is illustrated in Fig. \ref{f:basic}.

 The proposed \LOOPJCCP~ model is designed to yield an optimal solution to the problem, given input parameters and a set of uncertainty samples:

\begin{align}
    \mathbf{u}^{\triangle} = \xi_{\mathbf{w}}(\mathbf{x}, \{\boldsymbol{\epsilon}^{[k]}\}_{k=1}^{N_{\texttt{Scen}}}) \label{nn}
\end{align}
\normalsize

The model \( \xi_{\mathbf{w}} \) is developed with the following specific objectives:
\begin{enumerate}
    \item \underline{\textit{Near-optimality}}: The model's output should closely approximate the optimal solution of problem \eqref{power dispatch cc compact}.
    \item \underline{\textit{Feasibility Assurance}}: It should ensure that the constraints are met, adhering to a customizable probability setting of \( 1-\varepsilon \).
    \item \underline{\textit{Scalability and Flexibility}}: The model must accommodate different orders of input samples and be capable of handling varying numbers of samples, \( N_{\texttt{Scen}} \).
\end{enumerate}

Given the difficulties in integrating traditional reformulation methods into a neural network, our model utilizes the structure of \LOOPLCP~ \cite{li2023toward} and a novel polyhedron reformulation. Our model overcomes existing challenges and enhances performance in feasibility and adaptability.

\begin{figure*}[bp]
\centering
\setlength{\abovecaptionskip}{0.cm}
\includegraphics[width=2.05\columnwidth]{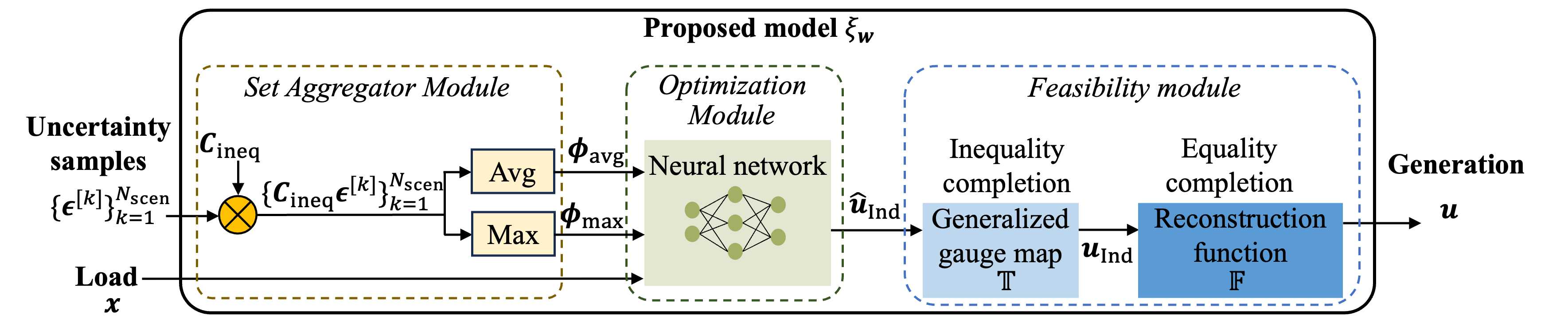}
\caption{Schematic of the proposed \LOOPJCCP~ model, which integrates the \LOOPLCP~ structure \cite{li2023toward} with a novel polyhedron reformulation. This model efficiently learns solutions to joint chance-constrained power dispatch problems within VPPs, delivering near-optimal approximations and ensuring feasibility with customizable probabilities. By incorporating XAI components, such as the set aggregate module and the closed-form feasibility module, it provides transparency and adaptability to diverse sample sizes and input sequences, enhancing interpretability and trust in the decision-making process.}
\centering
\label{f:overall}
\end{figure*}

As depicted in Fig. \ref{f:overall}, the proposed model comprises (i) the set aggregator module, (ii) the optimization module (a neural network), and (iii) the feasibility module. The following sections will detail different modules and their interconnections.

\subsection{Set Aggregator Module}
The set aggregator module is designed to provide a consistent and generalizable approach for dealing with variable sample sizes. This flexibility is crucial as the number of scenarios \( N_{\texttt{Scen}} \) may vary with different input parameters \( \mathbf{x} \). Additionally, the model's response should be permutation-invariant \cite{zaheer2017deep}, meaning it remains unaffected by the order of elements in the sample set. Formally, for any permutation \( \pi \), the model satisfies \( \xi_{\mathbf{w}}(\mathbf{x},[\boldsymbol{\epsilon}^{[1]},...,\boldsymbol{\epsilon}^{[N_{\texttt{Scen}}]}]) = \xi_{\mathbf{w}}(\mathbf{x},[\boldsymbol{\epsilon}^{[\pi(1)]},...,\boldsymbol{\epsilon}^{[\pi(N_{\texttt{Scen})]}}]) \).

To process the set input \( \{\boldsymbol{\epsilon}^{[k]}\}_{k=1}^{N_{\texttt{Scen}}}\), we employ aggregation methods using the average and maximum values as input features. Each element \( \boldsymbol{\epsilon}^{[k]} \) is first individually processed through the matrix \( \mathbf{C}_{\texttt{ineq}} \). The resulting vectors, the average \( \boldsymbol{\phi}_{\texttt{avg}} \) and the maximum \( \boldsymbol{\phi}_{\texttt{max}} \), are calculated as shown in \eqref{avg} and \eqref{max}. These aggregated representations are then combined with \( \mathbf{x} \) and fed into the optimization module.

\begin{gather}
    \boldsymbol{\phi}_{\texttt{avg}} = \frac{\sum_{k=1}^{N_{\texttt{Scen}}}\mathbf{C}_{\texttt{ineq}}\boldsymbol{\epsilon}^{[k]}}{N_{\texttt{Scen}}} \label{avg}\\
    \boldsymbol{\phi}_{\texttt{max}} = \max\left \{ \mathbf{C}_{\texttt{ineq}}\boldsymbol{\epsilon}^{[k]} \mid k = 1, \ldots, {N_{\texttt{Scen}}} \right \} \label{max}
\end{gather}
\normalsize

This module effectively handles varying numbers of samples in the joint chance-constrained power dispatch problem and ensures consistent outputs irrespective of input sequence order. The set aggregator module, implementable in code and compatible with backward propagation, can be integrated into various differentiable ML models.

The outputs of this module, the two single representations, are conveyed to the optimization module (a neural network) to derive a high-quality solution for problem \eqref{power dispatch cc}. Subsequently, the feasibility module verifies that the solution complies withconstraints, maintaining the adjustable probability threshold.

\subsection{Feasibility Module}
The feasibility module is crucial for ensuring solutions that are not overly conservative while still providing feasibility guarantees. These guarantees are probabilistic, ensuring compliance with the chance constraints. To achieve this, we introduce a new polyhedron reformulation, which effectively transforms the original joint chance-constrained problem into a deterministic format. This reformulation is designed for seamless integration into the ML model. Following this, we enhance the feasibility of the polyhedron reformulation by incorporating the \LOOPLCP~ model, as described in \cite{li2023toward}. The \LOOPLCP~ model employs techniques such as variable elimination and gauge mapping for managing equality and inequality constraints, respectively. These techniques enable the generation of solutions that are both feasible and close to optimal. By employing this closed-form, explainable feasibility module, our approach aligns with XAI principles, providing transparency and interpretability in how the model ensures constraint compliance. In subsequent sections, we delve into the specifics of the polyhedron reformulation and the integration of the \LOOPLCP~ model, detailing their roles in achieving the desired model outcomes.

\subsubsection{Proposed Polyhedron Reformulation}
In the feasibility module, we ensure that the model's outputs adhere to both equality constraints \eqref{power dispatch eq compact} and inequality constraints \eqref{eq:sa ineq para}:

\begin{align}
\mathbf{A}_{\texttt{ineq}}\mathbf{u} + \mathbf{B}_{\texttt{ineq}}\mathbf{x} + p\boldsymbol{\phi}_{\texttt{max}} + (1-p) \boldsymbol{\phi}_{\texttt{avg}} + \mathbf{b}_{\texttt{ineq}} \leq \mathbf{0} 
\label{eq:sa ineq para}
\end{align}
\normalsize

Here, \( p \) represents the safety parameter, ranging between [0,1]. Higher value of \( p \) increases the weight of $\boldsymbol{\phi}_{\texttt{max}}$, which results in a more conservative consideration of uncertainties. Differently put, a higher value of \( p \) directs the model to optimize within a smaller feasibility range. Specifically, at \( p = 1 \),  \eqref{eq:sa ineq para} simplifies to \eqref{eq:sa ineq max}. The above equation represents a compact and computationally efficient reformulation of the inequality constraints, which is particularly advantageous for handling large sample sets. Under this condition, the model optimizes similarly to the scenario approach, ensuring feasible solutions for every scenario in the sample set, often leading to lower than necessary violation rates.

\begin{align}
\mathbf{A}_{\texttt{ineq}}\mathbf{u}+\mathbf{B}_{\texttt{ineq}}\mathbf{x}+\max\{\mathbf{C}_{\texttt{ineq}}\boldsymbol{\epsilon}^k|k=1,...,N_{\texttt{Scen}}\}\nonumber\\
+\mathbf{b}_{\texttt{ineq}}\leq\textbf{0}
\label{eq:sa ineq max}
\end{align}
\normalsize

Conversely, with \( p \) approaching zero, the model considers a larger feasibility range with minimal uncertainty. Thus, as \( p \) varies from zero to one, the model's conservatism shifts from considering zero uncertainty to accommodating all uncertainty scenarios as feasible.

Therefore, by manipulating the safety parameter \( p \), the feasibility module can adjust the probability that the solution adheres to constraints \eqref{power dispatch eq compact} and \eqref{eq:sa ineq para}. The \LOOPJCCP~model is trained to optimize the problem as defined in \eqref{reform compact}, considering this adjustable parameter \( p \).

\begin{align}
   \min f(\mathbf{u}) \quad
\texttt{s.t.} \quad \mathbf{u} \in \mathcal{S} = \left \{\eqref{power dispatch eq compact}, \eqref{eq:sa ineq para} \right \} \label{reform compact}
\end{align}
\normalsize

\subsubsection{\LOOPLCP~ Model Integration}
To achieve hard feasibility with respect to the constraints of polyhedron reformulation, we integrate the \LOOPLCP~ model into the feasibility module. This module encompasses two submodules: equality completion and inequality completion, which will be discussed in detail.

\underline{\textit{Equality Completion Submodule}}: 
This submodule is designed to satisfy equality constraints \eqref{power dispatch eq compact} by eliminating variables. Given the equality constraints defined in \eqref{power dispatch eq compact}, we classify the variables \( \mathbf{u} \) into two groups: dependent variables \( \mathbf{u}_{\text{Dep}} \) and independent variables \( \mathbf{u}_{\text{Ind}} \). The transformation function \( \mathbb{F} \) maps \( \mathbf{u}_{\text{Ind}} \) to \( \mathbf{u}_{\text{Dep}} \), as outlined in the linear transformation suggested by \cite{li2023learning} and denoted in \eqref{eq:F}:

\begin{align}
    \mathbf{u}_{\texttt{Dep}} = \mathbb{F}(\mathbf{u}_{\texttt{Ind}})
    \label{eq:F}
\end{align}
\normalsize

Incorporating the relationship \( \mathbb{F} \) into \eqref{reform compact} and substituting \( \mathbf{u}_{\texttt{Dep}} \), we reformulate the feasible range in terms of \( \mathbf{u}_{\texttt{Ind}} \). The reformulated constraint set \( \mathcal{S}_{\texttt{Ref}} \) is given as \eqref{sref}.

\begin{align}
\mathcal{S}^{\texttt{Ref}}=\left \{ \mathbf{u}^{\texttt{Indep}}|\mathbf{A}_{\texttt{ineq}}\begin{bmatrix}
\mathbb{F}(\mathbf{u}_{\texttt{Ind}})\\ 
\mathbf{u}_{\texttt{Ind}}
\end{bmatrix}+\mathbf{B}_{\texttt{ineq}}\mathbf{x}+p\boldsymbol{\phi}_{\texttt{max}} \right.\nonumber\\
\left.+(1-p) \boldsymbol{\phi}_{\texttt{avg}}+\mathbf{b}_{\texttt{ineq}}\leq\textbf{0} \right \} 
\label{sref}
\end{align}
\normalsize

%The inequality completion module ensures that its output \( \mathbf{u}_{\texttt{Ind}} \) lies within \( \mathcal{S}_{\texttt{Ref}} \).

Given \( \mathbf{u}_{\texttt{Ind}} \in \mathcal{S}_{\texttt{Ref}}\), this submodule applies \( \mathbb{F} \) to generate a full-size \( \mathbf{u} = [\mathbf{u}_{\texttt{Dep}}, \mathbf{u}_{\texttt{Ind}}] \), ensuring compliance with \eqref{power dispatch eq compact}.

\underline{\textit{Inequality Completion Submodule}}: This submodule adheres to inequality constraints employing the generalized gauge map method from \cite{li2023toward}. This method maintains the feasibility of linear constraints by modifying predictions to fit within the desired range. The gauge map function, denoted as \( \mathbb{T} \), either maintains the predictions as-is if they are within the feasible range or scales them to the boundary, thus converting any infeasible predictions into feasible outputs \( \mathbf{u}_{\text{Ind}} \in \mathcal{S}_{\text{Ref}}\). The generalized gauge map \( \mathbb{T} \) is applied as per \eqref{eq:new_gauge_map}.

\begin{align}
    \mathbf{u}_{\texttt{Ind}} = \mathbb{T}(\mathbf{\hat{u}}_{\texttt{Ind}}) = \frac{\mathbf{\hat{u}}_{\texttt{Ind}}}{\max\left \{ 1, \psi_{\mathcal{S}_{\texttt{Ref0}}}(\mathbf{\hat{u}}_{\texttt{Ind}})\right \}} + \mathbf{u}_{\texttt{Ind,0}}
    \label{eq:new_gauge_map}
\end{align}
where $\mathbf{\hat{u}}_{\texttt{Ind}}$ is the direct prediction of the neural network with the same dimension as $\mathbf{u}_{\texttt{Ind}}$. 
$\mathcal{S}_{\texttt{Ref0}}=\left \{ \mathbf{\bar{u}}_{\texttt{Ind}}| \left (\mathbf{u}_{\texttt{Ind,0}}+\mathbf{\bar{u}}_{\texttt{Ind}}  \right )\in \mathcal{S}_{\texttt{Ref}}\right \}$ is a shifted set of $\mathcal{S}_{\texttt{Ref}}$ by
an interior point $\mathbf{u}_{\texttt{Ind,0}}$. $\psi_{\mathcal{S}_{\texttt{Ref0}}}(\mathbf{\hat{u}}_{\texttt{Ind}})$ is the Minkowski function value of $\mathbf{\hat{u}}_{\texttt{Ind}}$ on set ${\mathcal{S}_{\texttt{Ref0}}}$, defined as \eqref{psi_s}. 

\begin{align}
    \psi_{\mathcal{S}_{\texttt{Ref0}}}(\mathbf{\hat{u}}_{\texttt{Ind}})=\underset{r}{\max}\{\frac{\mathbf{A}^{\left \{ r \right \}}\mathbf{\hat{u}}_{\texttt{Ind}}}{{-\mathbf{A}^{\left \{ r \right \}}\mathbf{u}_{\texttt{Ind,0}}+\mathbf{b}(\mathbf{x})^{\left \{ r \right \}}}}\} \label{psi_s}
\end{align}

where $\mathbf{A}$ and $\mathbf{b}(\mathbf{x})$ constitute the compact parameters of $\mathcal{S}^{\texttt{Ref}}$, namely, equation \eqref{sref} can be condensed to $\mathcal{S}^{\texttt{Ref}}=\left \{ \mathbf{u}^{\texttt{Indep}}|\mathbf{A}\mathbf{u}_{\texttt{Ind}}\leq \mathbf{b}(\mathbf{x}) \right \}$. The superscript $r$ denotes the $r$th row in a vector (or matrix). 

Given any input prediction \( \mathbf{\hat{u}}_{\texttt{Ind}} \) from the neural network, this submodule ensures that \( \mathbf{u}_{\texttt{Ind}} \) remains within \( \mathcal{S}_{\texttt{Ref}} \).

In summary, with input parameter \( \mathbf{x} \) and set input \( \{\boldsymbol{\epsilon}^{[k]}\}_{k=1}^{N_{\texttt{Scen}}}\), the model produces a prediction \( \mathbf{u} \in \mathcal{S} \). Adjusting the parameter \( p \) alters \( \mathcal{S} \)'s range, ensuring that the prediction \( \mathbf{u} \) satisfies constraints with a customizable probability.

% \begin{table*}[htbp]
% \caption{Dataset Categories and usage: In-sample training, out-sample training, in-sample testing, and out-sample testing. }
% \begin{center}
% \label{table:dataset}
% \begin{tabular}{|c|cc|cc|}
% \hline
%                              & \multicolumn{2}{c|}{Offline training set}                        & \multicolumn{2}{c|}{Online testing set}                   \\ \hline
% Method                       & \multicolumn{1}{c|}{In-sample}            & Out-sample           & \multicolumn{1}{c|}{In-sample}         & Out-sample       \\ \hline
% \rowcolor{lighterorange}Scenario Approach            & \multicolumn{1}{c|}{-}                    & -                    & \multicolumn{1}{c|}{Model development} & Model validation \\ \hline
% \rowcolor{lighterorange}CVaR Method                  & \multicolumn{1}{c|}{-}                    & -                    & \multicolumn{1}{c|}{Model development} & Model validation \\ \hline
% \rowcolor{lighterorange}Robust Optimization          & \multicolumn{2}{c|}{Safe parameter selection}                    & \multicolumn{1}{c|}{Model development} & Model validation \\ \hline
% \rowcolor{lighterorange}Polyhedron Reformulation     & \multicolumn{2}{c|}{Safe parameter selection}                    & \multicolumn{1}{c|}{Model development} & Model validation \\ \hline
% \rowcolor{lightergreen} Proposed \LOOPJCCP~model & \multicolumn{2}{c|}{Model development, Safe parameter selection} & \multicolumn{1}{c|}{-}                 & Model validation \\ \hline 
% \end{tabular}
% \end{center}
% \vspace{-0.2in}
% \end{table*}

\subsection{Model Training}
\subsubsection{Dataset}
The datasets used in this study are categorized into four types: in-sample training, out-sample training, in-sample testing, and out-sample testing datasets.
Classical solver-based methods like the scenario approach and the CVaR method, which do not require parameter tuning, utilize only testing datasets. The in-sample testing data helps refine the model to discern underlying patterns, while the out-sample testing data evaluates the model's robustness on novel scenarios. For methods like robust and polyhedron reformulations that involve tuning a safety parameter, this parameter is selected offline using both in-sample and out-sample training datasets.

\begin{figure}[htbp]
\centering
\setlength{\abovecaptionskip}{0.cm}
\includegraphics[width=1\columnwidth]{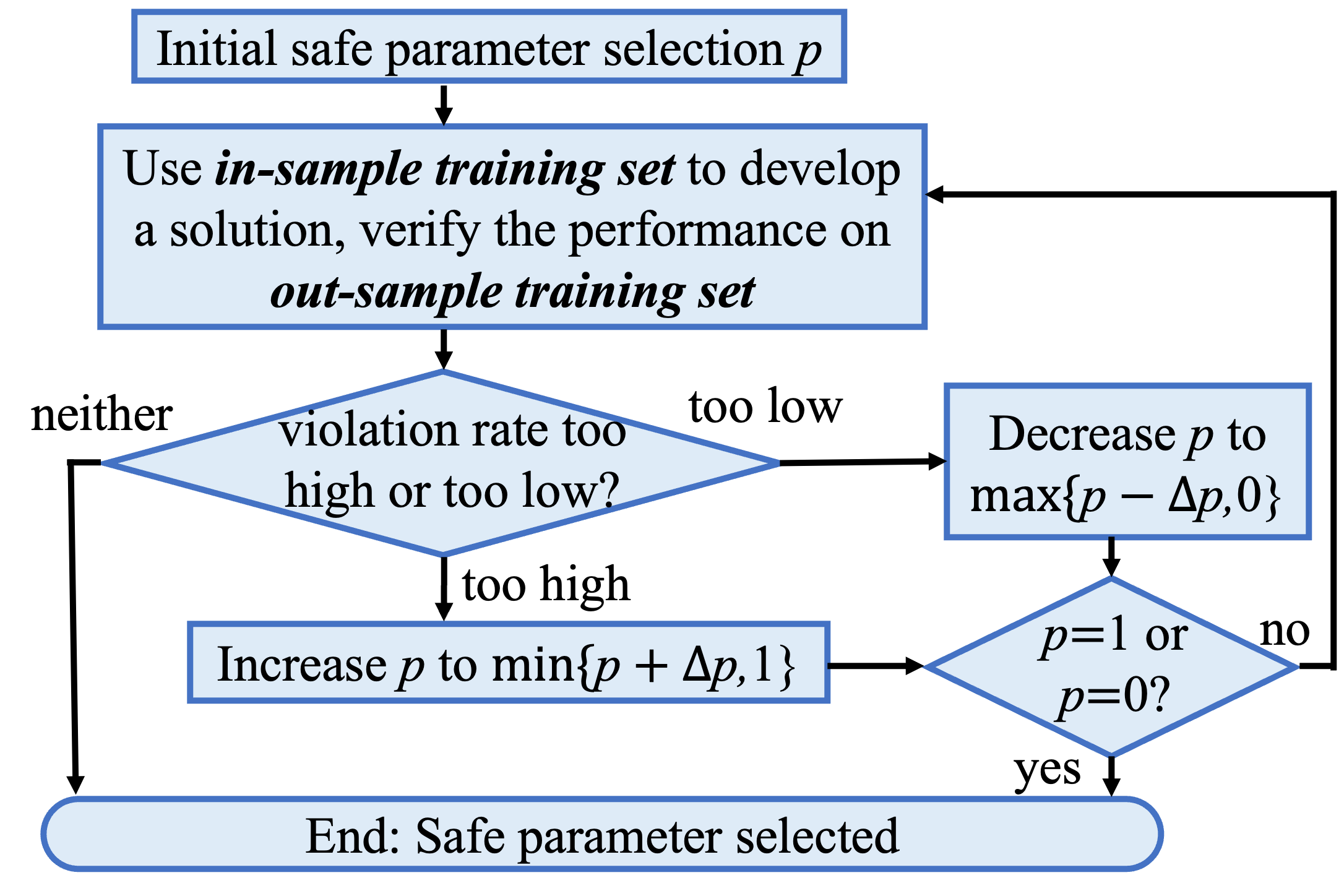}
\caption{Flow chart of safe parameter selection for the proposed model.}\centering
\label{flow}
\end{figure}

Our proposed ML-based model undergoes initial training with the in-sample training dataset, with performance validation and safety parameter selection using the out-sample training dataset, as shown in Fig. \ref{flow}. Final testing occurs online with the out-sample testing datasets to confirm the model's efficacy and performance.

\subsubsection{Loss function}

Our model allows two distinct training methodologies as outlined in \cite{liu2022teaching}: 1)
 Training \textit{with} a solver in the loop. 2) Training \textit{without} a solver in the loop, which involves direct minimization of the objective function.

In this research, we use the distance between the model's prediction, denoted as $\mathbf{u}^\triangle$, and the optimal solution obtained through polyhedron reformulation (as shown in \eqref{reform compact}). The optimal solution via polyhedron reformulation, represented as $\mathbf{u}^*_{\texttt{PR}}$, is computed using commercial solvers. This distance forms the basis of our loss function $L$, detailed in  \eqref{loss}.

\begin{align}
    L=\frac{1}{N}\sum_{n=1}^{N} \left \| \mathbf{u}^{(n)\triangle},\mathbf{u}^{(n)*}_{\texttt{PR}} \right \|^2_2\label{loss}
\end{align}
\normalsize

\noindent where $N$ denotes the number of training data points, and $(n)$ denotes its index.

\section{Experimental Results}
\subsection{Experiment Setup}
\subsubsection{Test Systems}
We analyze a Virtual Power Plant (VPP) consisting of 50 prosumers, half equipped with photovoltaic arrays and half with wind turbines. {\color{black}We employ the same coefficients for generators and loads as those used in \cite{li2023machine} and \cite{pan2015fractional}. The dispatchable generators operate within a range of $[0,80]$ kW, and flexible loads vary from $[10,25]$ kW, maintaining total output within $100$ kW.} % $\beta_{\texttt{G1}}^{i}=0.2$, $\beta_{\texttt{G2}}^{i}=0.01$, $\beta_{\texttt{L1}}^{i}=0.1$, and $\beta_{\texttt{L2}}^{i}=-3$, for $i = 1, \ldots, N_{\texttt{A}}$. 
Load profiles, influenced by data from central New York on July 24th, 2023 \cite{nyiso}, include a 10\% random fluctuation to differentiate individual prosumer profiles. Non-dispatchable generation from photovoltaic arrays spans $0-50$ kW, calculated using regional solar radiation intensity data from the Global CMP22 dataset on the same date \cite{stoffel1981nrel} and additional parameters from \cite{li2023machine}. Wind speeds from New York City's Central Park, also on July 24th, 2023 \cite{nws2023}, help estimate wind generation for each prosumer, ranging from $5-50$ kW.

\subsubsection{Training Data}
{\color{black}
To mitigate the risk of data inadequacies, we employed traditional solvers in this work to generate datasets for offline training.} We consider a fixed in-sample numbers of 1000 and out-sample number of 10000 for each data point.  We use a train/test ratio of 1:1 where odd indexed data points are for training and even indexed data points are for testing. We assume that all non-dispatchable generators fluctuate around their forecasted power with a standard deviation of 10\%.
We set the desired violation probability of the chance constraint to $\varepsilon = 0.05$, which is a value commonly used in power systems applications \cite{hou2022data}.

\subsubsection{Neural Network and Solver Configuration}

Our neural network models are designed with a single hidden layer, comprising 200 hidden units. The Rectified Linear Unit (ReLU) activation function is employed to introduce non-linearity into the models. For optimization computations in solver-based methods, we utilize the widely-accepted commercial solver, Gurobi \cite{gurobi}.

% For the training dataset, this value is $f(\mathbf{u}^*_{\texttt{SA}}) = 12715.8918$, and for the testing dataset, it is $f(\mathbf{u}^*_{\texttt{SA}}) = 12666.6004$. 
\subsubsection{Metrics for Comparative Analysis}

Our analysis employs two primary metrics for a comprehensive evaluation of the methods under study, the coresponding solution obtained using the considered method is denoted as $\mathbf{u}^*_{\text{method}}$. We run 10 times to use the average as the results.

$\bullet $ \textit{Objective Cost Rate}: This metric is designed to enable standardized comparisons across various methods. The objective cost rate is calculated by normalizing the objective function value of each method, $f(\mathbf{u}^*_{\texttt{method}})$ against a reference value. Specifically, this reference is the minimized objective function value obtained using the scenario approach, denoted as $f(\mathbf{u}^*_{\texttt{SA}})$. 
    The objective cost rate for the scenario approach itself is set to 1, establishing a baseline for comparison.

$\bullet $ \textit{In-sample and Out-sample Violation Rates}: This metric is determined by dividing the number of samples in which the solution $\mathbf{u}^*_{\text{method}}$ violates the joint chance constraint by the total number of samples.

% Formally, the objective cost rate for a given method is defined as:
% \begin{equation}
%     \text{objective cost rate} = \frac{V_{\text{method}}}{V_{\text{min,SA}}}
% \end{equation}
% where $V_{\text{method}}$ represents the objective function value for the method under consideration. 

\subsection{Offline Training Results}

\begin{figure*}[htbp]
\centering
\setlength{\abovecaptionskip}{0.cm}
\includegraphics[width=2\columnwidth]{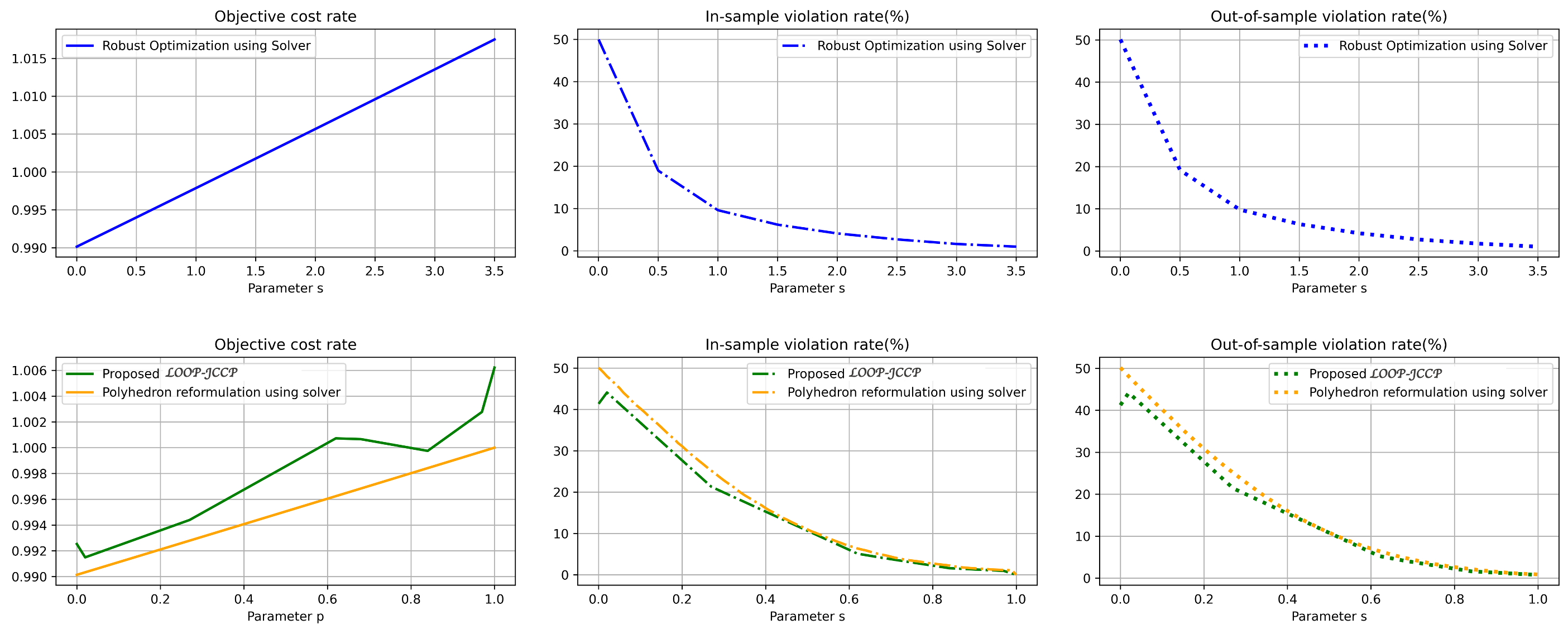}
\caption{Comparative analysis of robust optimization, polyhedron reformulation, and \LOOPJCCP~ using varying safety parameters to evaluate objective cost rate, in-sample, and out-sample violation rates. The objective cost rates are normalized against the scenario approach, expressed as  $f(\mathbf{u}^*_{\texttt{method}})/f(\mathbf{u}^*_{\texttt{SA}})$.
}
\centering
\label{f:train}
\end{figure*}

This section evaluates the impact of the safety parameter on the performance of the robust optimization method, polyhedron reformulation, and the proposed \LOOPJCCP~model using the training dataset. An optimal safety parameter is determined based on training performance and then applied in the online testing phase. It's important to note that the scenario approach and the CVaR method do not require parameter tuning and are not discussed here.

Fig. \ref{f:train} shows how performance varies with the safety parameter; as it increases, costs rise and conservatism grows, evidenced by lower violation rates. When the safety parameter is minimal, costs are below those of the scenario approach, indicating efficiency. At zero, the robust optimization and polyhedron reformulation methods treat uncertainties uniformly, based solely on sample set averages.

The \LOOPJCCP~model employs a generalized gauge map to ensure solutions stay within feasible limits, resulting in higher conservatism than the polyhedron reformulation when using the same safety parameter. The safety parameter range for \LOOPJCCP~is restricted to $[0, 1]$, simplifying selection compared to the robust optimization method, which has a more complex, open-ended range. This difference results in a nonlinear relationship between safety parameters and violation rates in robust optimization, complicating the targeted safety parameter selection.

% The implementation of the generalized gauge map in the proposed \LOOPJCCP~model confines its search within a feasible range. Consequently, when employing the same safety parameter $p$, the proposed \LOOPJCCP~model consistently demonstrates higher conservatism compared to the polyhedron reformulation, as evidenced by a higher objective cost rate and a lower violation rate.

% Regarding the selection of the safety parameter, the robust optimization method's parameter range $s$ extends beyond zero and is open-ended, making it more challenging to determine the appropriate value. In contrast, the safety parameter range $p$ for the proposed \LOOPJCCP~model is confined within [0, 1]. Furthermore, the relationship between the safety parameter and violation rate in the robust optimization method is more nonlinear than that in the proposed  \LOOPJCCP~model. This nonlinearity implies that identifying an appropriate safety parameter to achieve a targeted violation rate is more complex for the robust optimization method, as its violation rate changes unevenly with the safety parameter.

\subsection{Online Test Results}
After offline training, we selected safety parameters $s=1.98$ for the robust optimization method and $p=0.68$ for the proposed \LOOPJCCP~model. Table \ref{test results} shows the performance comparison on the testing dataset.

The CVaR method outperforms the scenario approach in objective cost rate, mainly because it focuses on the average outcomes of the worst cases, allowing for some infeasibility, unlike the scenario approach that requires feasibility in every scenario. Both methods, however, lack the flexibility to adjust violation rates due to their fixed parameters, leading to consistent performance with a given sample set.

Contrary to expectations, increased conservatism in the robust optimization method at $s=1.98$ led to higher costs and violation rates than the scenario approach. The proposed polyhedron reformulation, however, achieved the best objective cost rate, significantly aiding the \LOOPJCCP~model, which emulates this reformulation. While the \LOOPJCCP~model has a slightly higher objective cost rate compared to other solver-based methods, its faster execution time, detailed in Table \ref{test results}, provides a crucial advantage for VPPs that demand quick responses. {\color{black} The training is conducted off-line and on average the training time for our method takes between 1 to 2 seconds.} This efficiency is essential, especially when online test results require immediate corrective actions such as parameter retuning or model redevelopment.

\begin{table*}[tp]
\begin{center}
\caption{The performance of different methods  on testing data set}
\begin{tabular}{|c|c|c|c|c|c|}
\hline
Method                                              & Objective cost rate & In-sample violation rate(\%) & Out-sample violation rate(\%)  & Time(s) & Our time improvement 
%\LOOPJCCP's speed up
\\ \hline
\rowcolor{lighterorange}Scenario   Approach                                 & 1.0000              & 0.0000                   & 0.0087                      & 162.3415  & 48000x\\ \hline
\rowcolor{lighterorange}CVaR Method                                         & 0.9979              & 2.0903                   & 2.3368                      & 1194.4189 & 350000x\\ \hline
\rowcolor{lighterorange}Robust Optimization                                 & 1.0037              & 4.8750                   & 4.8528                      & 6.7044    &2000x\\ \hline
\rowcolor{lighterorange}Polyhedron Reformulation                            & 0.9967              & 4.8402                   & 4.5965                      & 6.1749   &1800x \\ \hline
\rowcolor{lightergreen}\LOOPJCCP~Model & 1.1011              & 4.4791                   & 4.3666                      & 0.0034   &- \\ \hline
\end{tabular}
\label{test results}
\end{center}
\end{table*}

\begin{figure}[htbp]
\centering
\setlength{\abovecaptionskip}{0.cm}
\includegraphics[width=0.85\columnwidth]{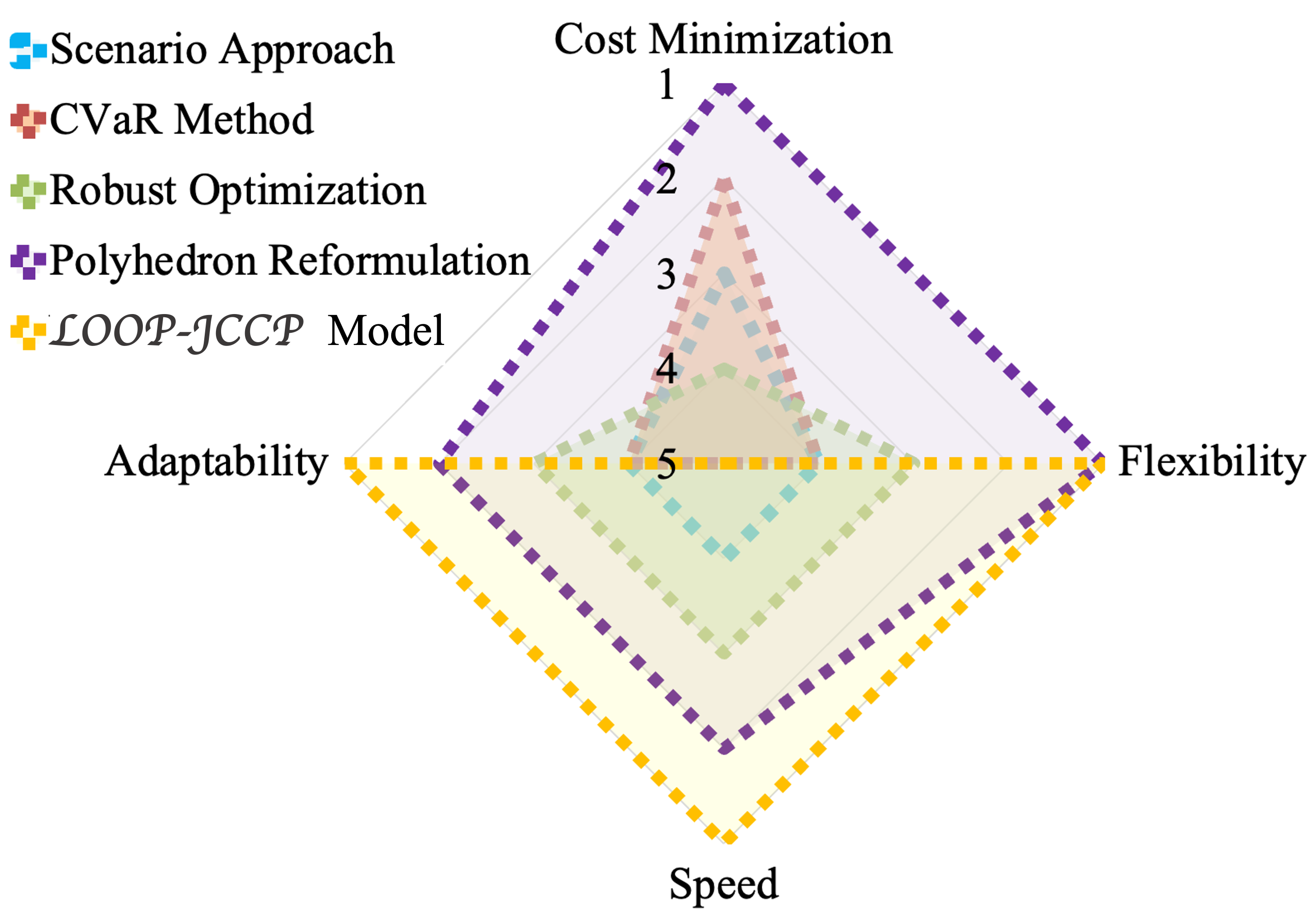}
\caption{We evaluate four key characteristics of different methods: \textit{Cost Minimization}, measuring the efficiency in reducing the objective cost rate; \textit{Flexibility to Meet Various Customized Violation Rates}, assessing the ability to adjust parameters for different violation rate requirements; \textit{Speed}, focusing on the computational efficiency and execution time of each method; and \textit{Adaptability to Different Testing Datasets}, gauging the responsiveness and flexibility of a method when faced with new or varied data conditions. 
}\centering
\label{f:radar}
\end{figure}

Fig. \ref{f:radar} provides a comprehensive summary of the characteristics of different methods applied to the same sample set in our experiment. Among these, the polyhedron reformulation stands out by achieving the lowest objective cost rate, thereby ranking highest in terms of cost minimization. When considering the flexibility to meet various customized violation rates, both the scenario approach and the CVaR method are less favorable. Their performances are fixed since there is no parameter available for tuning the violation rates, resulting in their lower ranking in this aspect. In terms of parameter range and ease of tuning, the robust optimization method is hindered by shortcomings. Its safety parameter range $s$ is larger than zero and open-ended, and exhibits a higher degree of nonlinearity, making it more challenging to determine the appropriate value. Conversely, the safety parameter range $p$ for the  \LOOPJCCP~model is confined within $[0,1]$, displaying a more linear relationship where the violation rate changes evenly with the safety parameter.

Regarding execution time,  \LOOPJCCP~model has the fastest performance compared to the other methods. In terms of adaptability to different testing datasets, the scenario approach and the CVaR method have the lowest rank due to their lack of adjustability. The robust optimization method is more adaptable than these two, but it is still behind the proposed  \LOOPJCCP~model. This is because the solver in the robust optimization method takes more time to update the safety parameter $s$ and to redevelop models. On the other hand, the proposed \LOOPJCCP~model demonstrates superior adaptability and responsiveness to changes in test data.

\section{Conclusion}
This paper addresses the significant challenges posed by integrating renewable energy sources into power systems, focusing on the substantial uncertainties involved. We developed \LOOPJCCP, a novel neural approximator, to predict optimal solutions for joint chance-constrained power dispatch problems, offering a faster and more flexible solution for managing these uncertainties. Our method features an innovative polyhedron reformulation, enabling the model to learn and adhere to specific constraints effectively.

By integrating explainable components, such as the set aggregate module and the closed-form feasibility module,  \LOOPJCCP~ aligns with XAI principles. This integration provides transparency and interpretability, allowing operators to understand, validate, and trust the solutions. Our model's parameter adjustment capabilities, adaptability to diverse data conditions, and significant reduction in execution time make it an invaluable tool for reliable power system operations, enhancing informed and transparent decision-making amidst the unpredictability of renewable energy sources.

\color{black}
\appendices
\section{ }

This section presents the detailed models of four classes of VPP assets that are discussed in Section II A in an abstract manner. These asset classes include inflexible loads, flexible loads,
dispatchable generators, and non-dispatchable generators. In what follows, we will provide a detailed model for each asset class \cite{li2023machine}.  

\subsubsection{Inflexible Loads} The inflexible loads are modeled as inputs to the decision-making problem and are represented by fixed quantities.
\subsubsection{Flexible Loads -- Heating, Ventilation, and Air Conditioning Systems}
We consider the following representation for the inverter-based heating, ventilation, and air conditioning systems. The power demand is denoted as $P_{\texttt{HVAC}}^{i,t}$.

\begin{align}
T_{\texttt{HVAC}}^{i,t}\!\!=\!\varepsilon^i_{\texttt{HVAC}} T_{\texttt{HVAC}}^{i,t-1}\!\!+\!(1\!\!-\!\!\varepsilon^i_{\texttt{HVAC}} )\left (  T_{\texttt{out}}^{i,t-1}\!\!-\!\!\frac{\eta_{\texttt{HVAC}}^i }{A_{\texttt{HVAC}}^i}P_{\texttt{HVAC}}^{i,t}\right )\label{HVAC time step} 
\end{align}
\normalsize

here $T_{\texttt{HVAC}}^{i,t}$ is the indoor temperature, $T_{\texttt{out}}^{i,t}$ indicates the forecasted outdoor temperature, $\varepsilon^i_{\texttt{HVAC}}$ is the inertia factor, $\eta_{\texttt{HVAC}}^i$ is the coefficient of performance, $A_{\texttt{HVAC}}^i$ is thermal conductivity. The comfort can be adjusted within a range i.e., $[T^i_{\texttt{min}},T^i_{\texttt{max}}]$ (see \eqref{HVAC temperature}). Also, \eqref{HVAC power} limits the control to the size of air-conditioning (i.e., $P^{i}_{\texttt{HVACmax}}$).

\begin{align}
& T^i_{\texttt{min}}\leq T_{\texttt{HVAC}}^{i,t}\leq T^i_{\texttt{max}}\label{HVAC temperature}\\
&0\leq P_{\texttt{HVAC}}^{i,t}\leq P^{i}_{\texttt{HVACmax}}\label{HVAC power}
\end{align}
%T^i_{\texttt{HVAC}}\tau is known, T_{\texttt{HVAC}}^{i,t+1} is dependent variable
\normalsize

In this paper, we consider a single time step; hence, the coupling constraints can be represented by up-lower bounds.

The HVAC temperature adjustments should not adversely impact building occupants. To this end, \(\alpha_{\texttt{HVAC}}^{i,t}\) captures the cost coefficient, \(T_{\texttt{Ref}}^{i,t}\) defines the optimal comfort temperature, and binary variable \(\beta_{\texttt{HVAC}}^{i,t}\) denotes occupancy state, with zero indicating vacancy and one otherwise.

\begin{align}
f_{\texttt{HVAC}}=\sum_{t}\sum_{i\in \mathcal{N}_{\texttt{A}}}   \beta_{\texttt{HVAC}}^{i,t} \alpha_{\texttt{HVAC}}^{i}{(T_{\texttt{HVAC}}^{i,t}- T_{\texttt{Ref}}^{{i,t}}  )}^2
\end{align}
\normalsize

\subsubsection{Flexible Loads -- Plug-in Electric Vehicles (PEV)}

The PEV charging power \(P_{\texttt{PEV}}^{i,t}\) must respect the range \([P_{\texttt{PEVmin}}^{i},P_{\texttt{PEVmax}}^{i}]\) as described in \eqref{PEV power}. Further,  \eqref{PEV energy} ensures that prosumer \(i\)'s cumulative charging demand meets the necessary energy \(E_{\texttt{PEV}}^{i}\) for daily commute.

\begin{align}
P_{\texttt{PEVmin}}^{i}\leq P_{\texttt{PEV}}^{i,t}\leq P_{\texttt{PEVmax}}^{i}\label{PEV power}\\
   \sum_{t} P_{\texttt{PEV}}^{i,t}\geq E_{\texttt{PEV}}^{i}
 \label{PEV energy}
\end{align}
\normalsize

\subsubsection{Other Flexible Loads}
  At any time $t$, the power of a flexible load
 should be within a pre-defined operation range $[P_{\texttt{FLmin}}^{i,t},P_{\texttt{FLmax}}^{i,t}]$,  for any prosumer $i\in \mathcal{N}_{\texttt{A}}$:

\begin{align}
P_{\texttt{FLmin}}^{i,t} \leq  P_{\texttt{FL}}^{i,t}  \leq  P_{\texttt{FLmax}}^{i,t} 
\label{flexible load}
\end{align}
\normalsize

The objective is to
limit the differences between the optimized and baseline consumption profiles (i.e., \(P_{\texttt{FLref}}^{i,t}\)) for flexible loads. Here, \(\alpha_{\texttt{FL}}^{i,t}\) is the inconvenience coefficient. %, and \(P_{\texttt{FLref}}^{i,t}\) specifies the preferred consumption level.

\begin{align}
f_{\texttt{FL}}=\sum_{t}\sum_{i\in \mathcal{N}_{\texttt{A}}}  
\alpha_{\texttt{FL}}^{i}( P_{\texttt{FL}}^{i,t}- P_{\texttt{FLref}}^{i,t} )^2
\end{align}
\normalsize

\subsubsection{Dispatchable generator and flexible load -- Energy Storage Systems}

Depending on energy storage's charging and discharging status it can be classified as a dispatchable generator or a flexible load. 
 
 At any time $t$, the charging \(P_{\texttt{ESSC}}^{i,t}\) (or discharging \(P_{\texttt{ESSD}}^{i,t}\)) power of the energy storage system must not exceed \(P_{\texttt{ESSmax}}^{i}\), as indicated in \eqref{ess charging_discharging limit}. Also, \eqref{ess soc def} and \eqref{ess soc limit} define $R_{\texttt{SoC}}^{i,t}$ as the state of charge (SoC) and bound its limits. Here $\eta _{\texttt{ESSC}}^{i}$ and $\eta _{\texttt{ESSD}}^{i}$ denote the charging and discharging efficiencies. Finally $E_{\texttt{ESSN}}^{i}$ refers to the capacity.

\begin{align}
0\leq P_{\texttt{ESSC}}^{i,t}&\leq P_{\texttt{ESSmax}}^{i}, ~~~
0\leq P_{\texttt{ESSD}}^{i,t}\leq P_{\texttt{ESSmax}}^{i}  \label{ess charging_discharging limit}\\
R_{\texttt{SoC}}^{i,t+1}=&R_{\texttt{SoC}}^{i,t}+\frac{(P_{\texttt{ESSC}}^{i,t}\eta _{\texttt{ESSC}}^{i}- \frac{P_{\texttt{ESSD}}^{i,t}}{\eta _{\texttt{ESSD}}^{i}})\Delta t }{E_{\texttt{ESSN}}^{i}}
\label{ess soc def}
\\
&R_{\texttt{SoCmin}}^{i}\!\!\leq R_{\texttt{SoC}}^{i,t+1} \leq\!\! R_{\texttt{SoCmax}}^{i}
\label{ess soc limit}
\end{align}
\normalsize

The energy storage systems aim to simultaneously minimize maintenance and operation costs as shown in \eqref{objective function ess}. The coefficient \(\alpha_{\texttt{ESS}}^{i}\) connects the maintenance cost with charging and discharging behavior.

\begin{align}
f_{\texttt{ESS}}=\sum_{t} \sum_{i\in \mathcal{N}_{\texttt{A}}}   \alpha_{\texttt{ESS}}^{i}(  P_{\texttt{ESSC}}^{i,t}+P_{\texttt{ESSD}}^{i,t} ) \label{objective function ess}
\end{align}
\normalsize
% one time slot, up-lower bound

% simple battery \cite{he2019distributed} ??????

%R_{\texttt{SoC}}^{i,\tau} kniwn, to detemine P_t, R_{t+1} is dependent variable

\subsubsection{Non-dispatchable generators -- Photovoltaic (PV) Arrays}
The PV power output, given by \eqref{PV}, is determined by the solar irradiance-power conversion function. Here, $R_{\texttt{PV}}^t$, represents the solar radiation intensity, $A_{\texttt{PV}}$ denotes the surface area, and $\eta_{\texttt{PV}}$ is the transformation efficiency.

 \begin{align}
P_{\texttt{PV}}^{i,t}= R_{\texttt{PV}}^t A_{\texttt{PV}} \eta_{\texttt{PV}}\label{PV}
\end{align}
\normalsize

\subsubsection{Non-dispatchable generators -- Wind Turbines}
% https://engineercalcs.com/how-to-calculate-wind-turbine-power-output/
Wind turbine plant generation is given by \cite{pan2015fractional}:

 \begin{align}
P_{\texttt{Wind}}^{i,t}= 1/2 \rho_{\texttt{Wind}} A_{\texttt{Wind}} C_{\texttt{Wind}}^{t} {V_{\texttt{Wind}}^{t}}^3 \label{wind}
\end{align}
\normalsize

\noindent where, $\rho_{\texttt{Wind}}$ denotes air density. Also, $ A_{\texttt{Wind}}$ shows the swept area of the blades, $ C_{\texttt{Wind}}^t$ denotes power coefficient, and $V_{\texttt{Wind}}^t$ captures the wind speed.
\color{black}
\section*{Acknowledgment}

We gratefully acknowledge Professor Daniel K. Molzahn for his contributions to the conceptual discussions and inspiration that significantly shaped this paper. This work is supported by the National Science Foundation under grant no. \# 2313768.

\bibliographystyle{ieeetr}
\bibliography{mainLatex-for-paper-recreation.bib}

\begin{IEEEbiography}[{\includegraphics[width=1in,height=1.25in,clip,keepaspectratio]{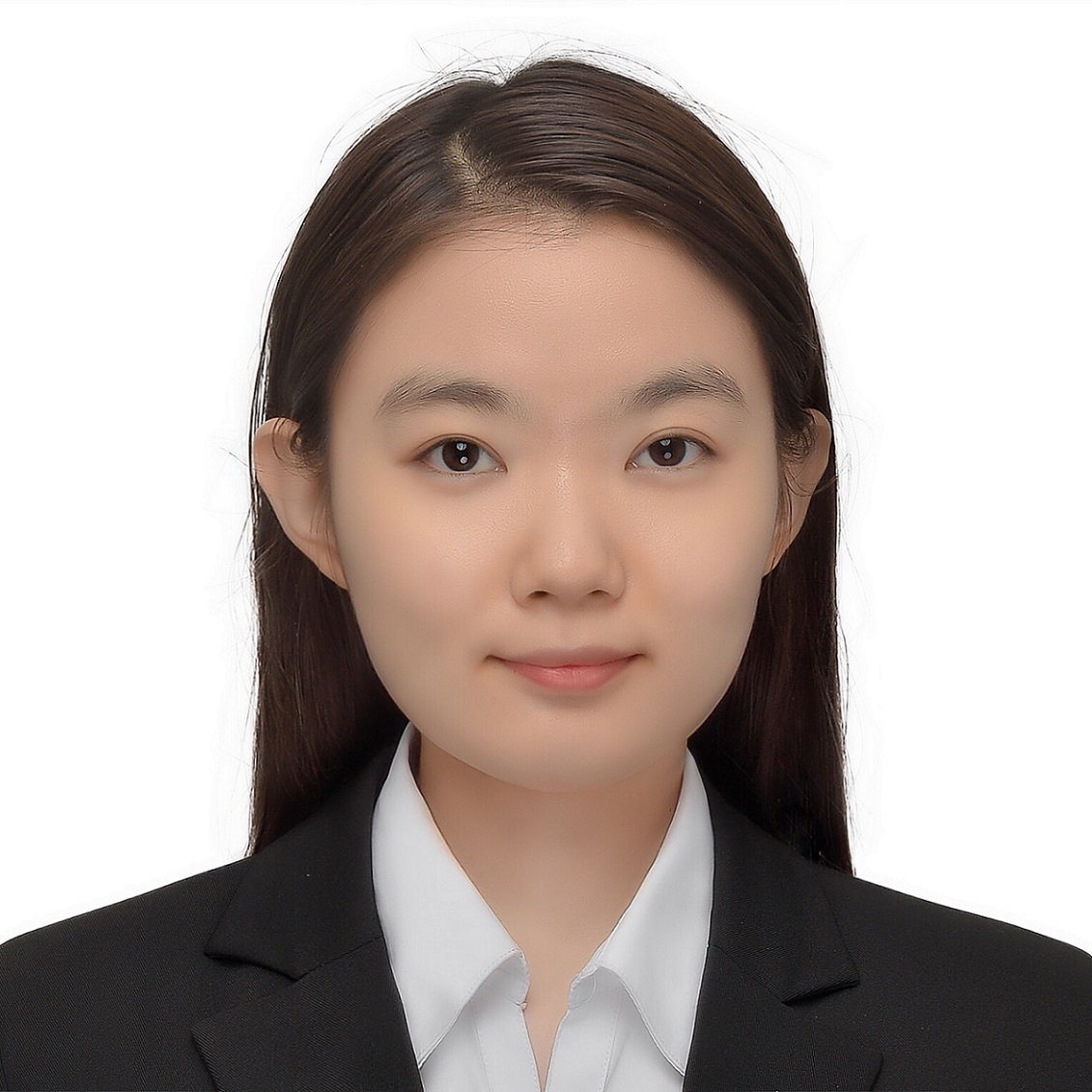}}]{Meiyi Li} obtained her B.S. and M.S. degrees in electrical engineering from Shanghai Jiao Tong University, Shanghai, China, in 2017 and 2020, respectively. She began her Ph.D. at Carnegie Mellon University, Pittsburgh, PA, USA, from 2020 to 2021 and has continued her doctoral studies at the University of Texas at Austin, Austin, TX, USA, since 2022. 

She has received numerous accolades, including the prestigious "Best of the Best" Paper Award at the 2019 IEEE Power and Energy Society General Meeting. Her impressive academic record includes the National Scholarship for Outstanding Academic Achievements in 2018, the Carnegie Institute of Technology Dean's Fellowship in 2020, and the Distinguished Alumni Endowed Graduate Fellowship at the University of Texas at Austin in 2022. Meiyi's research interests encompass learning to optimize and distributed optimization within the power and energy domain.
\end{IEEEbiography}

\begin{IEEEbiography}[{\includegraphics[width=1in,height=1.25in,clip,keepaspectratio]{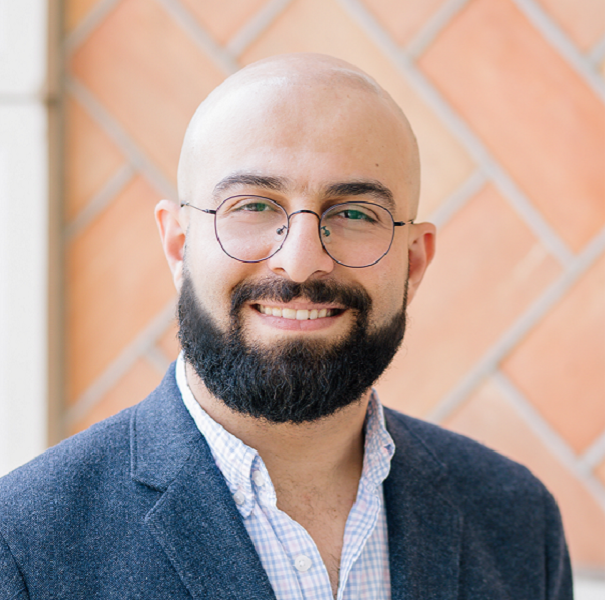}}]{Javad Mohammadi}  received his Ph.D. from the Electrical and Computer Engineering Department at Carnegie Mellon University (CMU) in 2016. As a graduate student, he was a recipient of the Innovation Fellowship from the Swartz Center for Entrepreneurship. Javad is an IEEE senior member.

He is currently an assistant professor in the Department of Civil, Architectural, and Environmental Engineering at The University of Texas at Austin. Before joining UT, he was a faculty member in CMU's Electrical and Computer Engineering department. His research interests include distributed decision-making in networked cyber-physical systems, including energy networks and electrified transportation systems. AFOSR, ARPA-E, and the Department of Energy support his grid modernization efforts. His research on building energy management has received financial support from local governments and institutions such as the Sloan Foundation.
\end{IEEEbiography}

\end{document}